\documentclass[rnaas, modern]{aastex63}

\usepackage{amsmath}

\shorttitle{Late-Time Light Curve of SN 1992A}
\shortauthors{Roche \&\ Garnavich}

\newcommand{\cosix}{$^{56}$Co}
\newcommand{\coseven}{$^{57}$Co}

\begin{document}

\title{Testing Progenitor Models Using the Late-Time Light Curve of Supernova 1992A}

\author{Cian Roche}
\affiliation{Department of Physics, Universit\"at T\"ubingen, 72074 T\"ubingen}
\affiliation{Naughton REU Fellow}
\author{Peter Garnavich}
\affiliation{Department of Physics, University of Notre Dame, Notre Dame, IN 46556}

\correspondingauthor{Peter Garnavich}
\email{pgarnavi@.nd.edu}


\begin{abstract}

The dominant radioactive energy source powering Type~Ia supernova light curves is expected to switch from the decay of \cosix\ to \coseven\ at very late epochs. We use archival $HST$ images of SN1992A obtained more than 900~days after explosion to constrain its cobalt isotopic abundance ratio and compare it to the well-studied event SN2011fe. We confirm the \coseven /\cosix\ ratio for SN2011fe of 0.026$\pm 0.004$ found by \citet{shappee17}, consistent with a ``double degenerate'' progenitor scenario. For SN1992A, we find a ratio of 0.034$\pm 0.010$, but the large uncertainty does not allow us to differentiate between progenitor models.

\end{abstract}

\section{Introduction}

Type~Ia supernovae (SNIa) are thermonuclear explosions of carbon-oxygen white dwarf (WD) stars. The explosions yield large amounts of radioactive nickel and the energy from radioactive decay powers their light curves. The progenitors of SNIa are believed to be binary systems, but their exact nature remains uncertain. The WD may reach instability by accreting matter from a nondegenerate secondary star (aka ``single degenerate" or SD model) \citep{whelan73}. Alternatively, two degenerate WDs in a tight binary may merge through the loss of energy and angular momentum by the emission of gravitational radiation (aka ``double degenerate"  or DD model)  \citep{1984ApJS...54..335I, 1984ApJ...277..355W}.

Observationally, it has been difficult to differentiate between the SD and DD scenarios. One possible probe is to determine isotopic abundances in the thermonuclear debris. \citet{ropke12} predicted that the nuclear burning for the SD channel occurs under denser conditions than in DD explosions. These dense conditions tend to yield an increased fraction of neutron-rich nickel, permitting a way to observationally differentiate between these progenitors. The radioactive nickel decays to cobalt and then to iron with varying half-lives:

\begin{centering}
$$
^{56}Ni \xrightarrow[]{t_{1/2} = 6.08d}\ ^{56}Co\   \xrightarrow[]{t_{1/2} = 77.2d}\ ^{56}Fe\ 
$$
$$
^{57}Ni \xrightarrow[]{t_{1/2} = 35.6h}\ ^{57}Co\   \xrightarrow[]{t_{1/2} = 271.79d}\ ^{57}Fe\ .
$$
\end{centering}

The \cosix\ isotope is expected to be between 30 and 40 times more abundant than \coseven , but its very long half-life means that \coseven\ should eventually dominate the luminosity of the supernova debris.

Here, we compare the late-time light curve of two well-studied events, SN2011fe and SN1992A, with the goal of constraining the abundance ratio between \coseven\ and \cosix .  Deep $HST$ imaging of SN2011fe out to 1800~days after peak was used by 
\citet{shappee17} to find a ratio consistent with the DD channel. In contrast, other studies concluded that there are major systematic uncertainties in the data that prevent an accurate estimate of the isotopic ratios in SN2011fe \citep{dimitriadis17, kerzendorf17, graur18}. We make a photometric measurement of SN1992A at an age of 900~days based on archival $HST$
imaging with the hope of constraining the \coseven/\cosix\ ratio in that explosion.

\begin{figure}
    \centering
    \includegraphics[width=\textwidth]{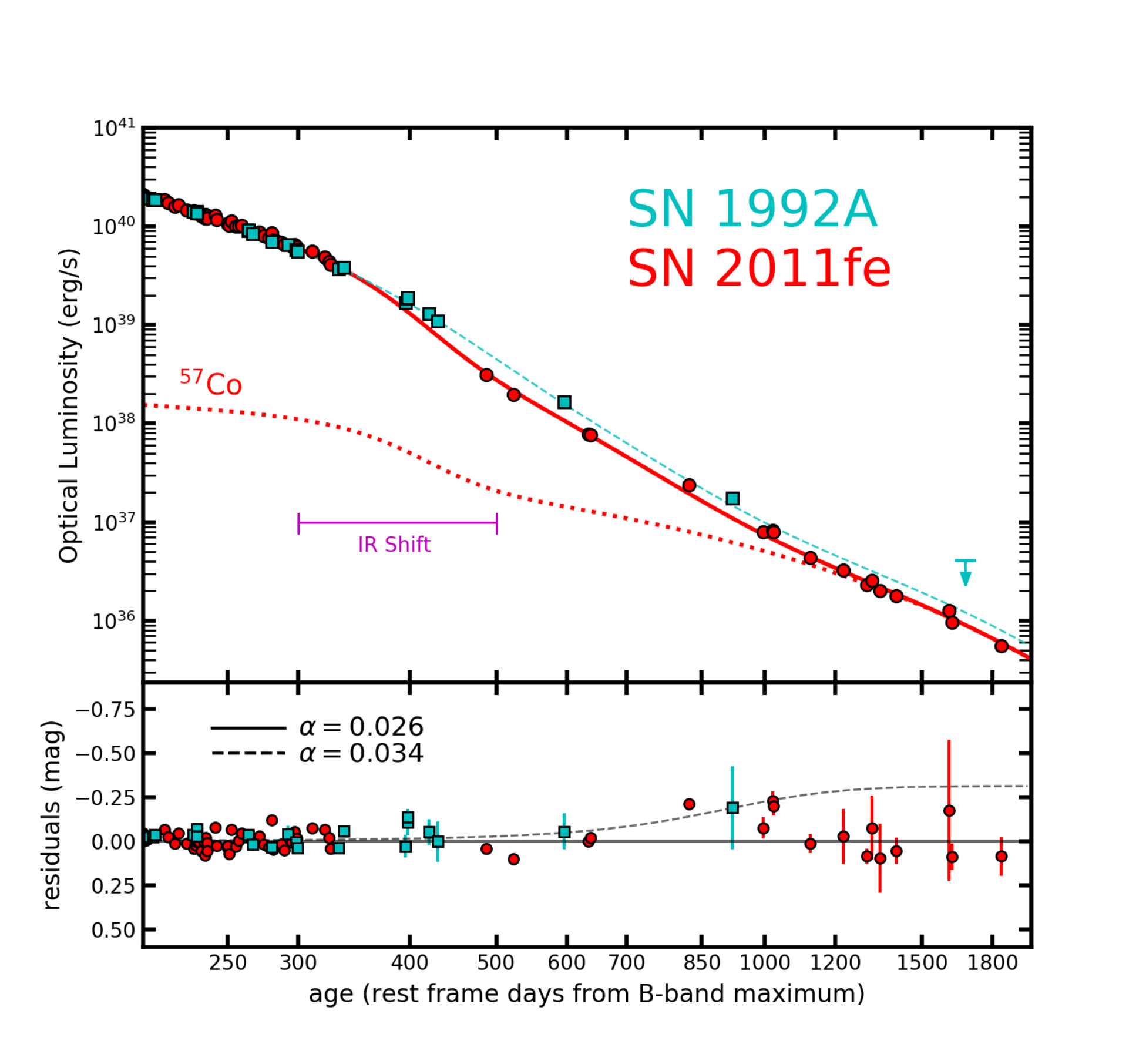}
    \caption{{\bf top:} The late light curves of SN1992A (squares) and SN2011fe (circles) created from the $B$ and $V$ filter observations. The dashed line indicates the best fit radioactive decay model for SN1992A while the solid line shows the same for SN2011fe. The dotted line indicates the contribution to the luminosity of SN2011fe from the decay of \coseven . The epoch for the shift of flux to the IR is indicated by the magenta bar {\bf bottom:} The residuals in magnitudes between the models and the observations. The best fit cobalt isotope ratio for SN2011fe is $\alpha =0.026\pm 0.004$. The ratio for SN1992A is based on the single observation at 900~days which gives $\alpha =0.034\pm 0.010$. The dashed line indicates a model containing a cobalt isotope ratio of 0.034. }
    \label{fig1}
\end{figure}

\section{Data}

We use the early light curve of SN2011fe from \citet{munari13} and the late-time photometry from \citet{shappee17} . The light curve of SN1992A was obtained from \citet{suntzeff}. 

SN1992A was observed with $HST/WFPC2$ on 1994 August~1,
926~days after $B$-band maximum light. Imaging was done in the $F439W$ and $F555W$ filters which are similar to the standard Johnson-Cousins $B$ and $V$ bandpasses. The field was deeply imaged again on 1996, September 4, but the supernova was no longer detectable. Aperture photometry was performed on the images and nearby stars were used to correct the small aperture to a 0.5~arcsec radius aperture \citep{holtzman95}. The count rate was then converted to standard $B$ and $V$ magnitudes using the prescription of \citet{holtzman95}. The uncertainties on the photometry as well as the limit on the non-detection were estimated by placing 100 apertures randomly about the host galaxy and determining their standard deviation. The resulting magnitudes for SN1992A at an age of 926~days are $B=26.06\pm 0.37$~mag and $V=25.78\pm 0.22$~mag. The 3$\sigma$ upper limit at an age of 1691~days was $V>27.36$~ mag.

Since the very late photometry of SN1992A were obtained only in the $B$ and $V$ filters, we created pseudo-bolometric light curves for SN1992A and SN2011fe over the optical wavelength range. This was accomplished by converting the $B$ and $V$ magnitudes in the Vega system to flux densities, fitting a line to the flux densities at their effective wavelengths, and then integrating from 3900~\AA\ to 6000~\AA . These wavelengths were chosen as they correspond to the blue and red edges of the two filters. Conveniently, these combined bands intercept the bulk of the flux from normal SNIa at late ages \citep[e.g.][]{mcclelland}. A year or so after explosion however, significant luminosity from SNIa shifts from the optical to the infrared (IR). This was
noted in SN2011fe and several other SNIa by \citet{dimitriadis17} and we use their sigmoid function to account for the decreased optical contribution.

The optical luminosity as a function of time is shown for both supernovae in Figure~\ref{fig1}. We have assumed a distance to NGC~1380, the host of SN1992A, of 18.70~Mpc. The precise distance is not critical to the analysis as it is degenerate with the mass of radioactive nickel synthesized in the explosion. The luminosity of SN2011fe has been adjusted to match that of SN1992A between 200 and 300 days and the observed ages are corrected to their rest frames using each host's heliocentric redshifts.

\section{Analysis and Conclusions}

We use the description of SNIa light curve physics provided by \citet{dimitriadis17} to compare with the observations. 
However, we assume that the optical depth time scale for gamma-rays is 35~days for both decay chains and we assume complete trapping of the lepton and x-ray energies  emitted in the decays. The model is a poor fit without accounting for the shift in optical luminosity to the IR. We multiply by a sigmoid function centered at an age of 400~days to model the shift. The amplitude of the luminosity shift is adjusted to match the observations between 300 and 600~days. As shown in the lower panel of Figure~\ref{fig1}, the SN2011fe model is best fit by decreasing the optical luminosity by a factor of 4.0. However, to fit the SN1992A light curve, a smaller IR shift factor of 2.9 better matches the data. The difference in the shifts is easily seen as SN1992A remains brighter than SN2011fe for ages later than 400 days. Clearly, there is not a single IR shift factor that will work for all supernovae, so this correction must be fit individually using the optical observations if no IR data is available. 

The light curve model beyond an age of 700~days is very sensitive to the isotopic mass abundance ratio, $\alpha =$\coseven/\cosix . For SN2011fe, the $\chi^2$ parameter is minimized for $\alpha =0.026\pm 0.004$. Despite a different approach, our value matches the results from \citet{shappee17}, who claimed that this $\alpha$ value is consistent with a low explosion density found in the DD scenario. Using the same technique, we find an isotopic ratio of $\alpha = 0.034\pm 0.010$ for SN1992A. Unfortunately, with only a single photometric detection beyond 700~days, the uncertainty in this measurement is too large to differentiate between SD and DD progenitors.  


\acknowledgments

We thank R. Kirshner, N. Suntzeff, B. Shappee, and the Naughton Fellowship REU Program.

{}

\end{document}